\documentclass[
,final            
  ]
  {aipproc}

\layoutstyle{6x9}

\usepackage{amssymb,amsmath}
\usepackage{subfigure}

\newcommand{\pip}{$\vec \gamma \vec p \rightarrow \pi^+n${ }}

\begin{document}

\title{Helicity Asymmetry $E$ in $\vec\gamma \vec p \to \pi^+n$ with FROST}


\author{Steffen Strauch for the CLAS Collaboration}{
  address={University of South Carolina, Columbia, SC 29208, USA},
  email={strauch@sc.edu}
}

\begin{abstract}
 The main objective of the FROST experiment at Jefferson Lab is the
study of baryon resonances. The polarization observable $E$ for the
reaction $\vec \gamma \vec p \to \pi^+n$ has been measured as part of
this program. A circularly polarized tagged photon beam with energies
from 0.35 to 2.35 GeV was incident on a longitudinally polarized
frozen-spin butanol target. The final-state pions were detected with
the CEBAF Large Acceptance Spectrometer.  Preliminary polarization
data agree fairly well with present SAID and MAID partial-wave
analyses at low photon energies. In most of the covered energy range,
however, significant deviations are observed. These discrepancies
underline the crucial importance of polarization observables to
further constrain these analyses.
\end{abstract}

\maketitle


\section{Introduction}

Pion photoproduction has long been analyzed in order to extract the
photodecay amplitudes associated with the $N^*$ and $\Delta^*$
resonances.  To unambiguously determine the four complex reaction
amplitudes a formally complete experiment requires the measurement of
at least eight observables at each energy and angle; four of which
need to be appropriately chosen double-spin observables
\cite{Chiang:1996em}.  However, the current database for pion
photoproduction is populated mainly by unpolarized cross sections and
single-spin observables.  The present work represents a measurement of
the double-polarization observable $E$ in the \pip reaction of
circularly polarized photons with longitudinally polarized
protons. The polarized cross section is in this case given by
\cite{Barker:1975bp}
\begin{equation}
  \frac{d\sigma}{d\Omega}=\frac{d\sigma}{d\Omega}_0 \left(1-P_zP_\odot E\right),
\end{equation}
where $d\sigma/d\Omega_0$ is the unpolarized cross section; $P_z$
and $P_\odot$ are degrees of target and beam polarizations, respectively.

\section{Experiment}

The double-polarization observable $E$ was studied in the \pip
reaction with the CEBAF Large Acceptance Spectrometer (CLAS)
\cite{Mecking:2003zu} at Thomas Jefferson National Accelerator
Facility.  Longitudinally polarized electrons with energies of
1.65~GeV and 2.48~GeV were incident on the thin radiator of the Hall-B
Photon Tagger \cite{Sober:2000we} and produced circularly-polarized
tagged photons in the energy range between 0.35~GeV and 2.35~GeV. The
circular polarization of the photon beam was determined from the
electron-beam polarization, $P_e \approx 85$\%, and the ratio of
photon and incident electron energy \cite{Olsen:1959zz}.  The
photon-helicity state changes with the electron-beam helicity, which
was flipped pseudorandomly at a rate of 30~Hz.  The electron
beam-charge asymmetry was negligible.  The collimated photon beam
irradiated the Hall B Frozen-Spin butanol target.  In this experiment
the proton spin was alined with the beam axis; the degree of proton
polarization was on average about 80\%. The target polarization
decreased over time with a time constant of the order of 100 days.
Typically, once per week the target was repolarized or the
polarization direction flipped.  Data were taken simultaneously on a
$^{12}$C target to allow for the study of background contributions
from photoproduction off bound nucleons in the butanol target.  The
$^{12}$C target was located a few centimeters downstream of the
butanol target.  The \pip reaction channel was identified by the
missing-mass technique with the $\pi^+$ meson detected in CLAS.

\section{Results}

\begin{figure}[htb!]
  \includegraphics[height=.3\textheight]{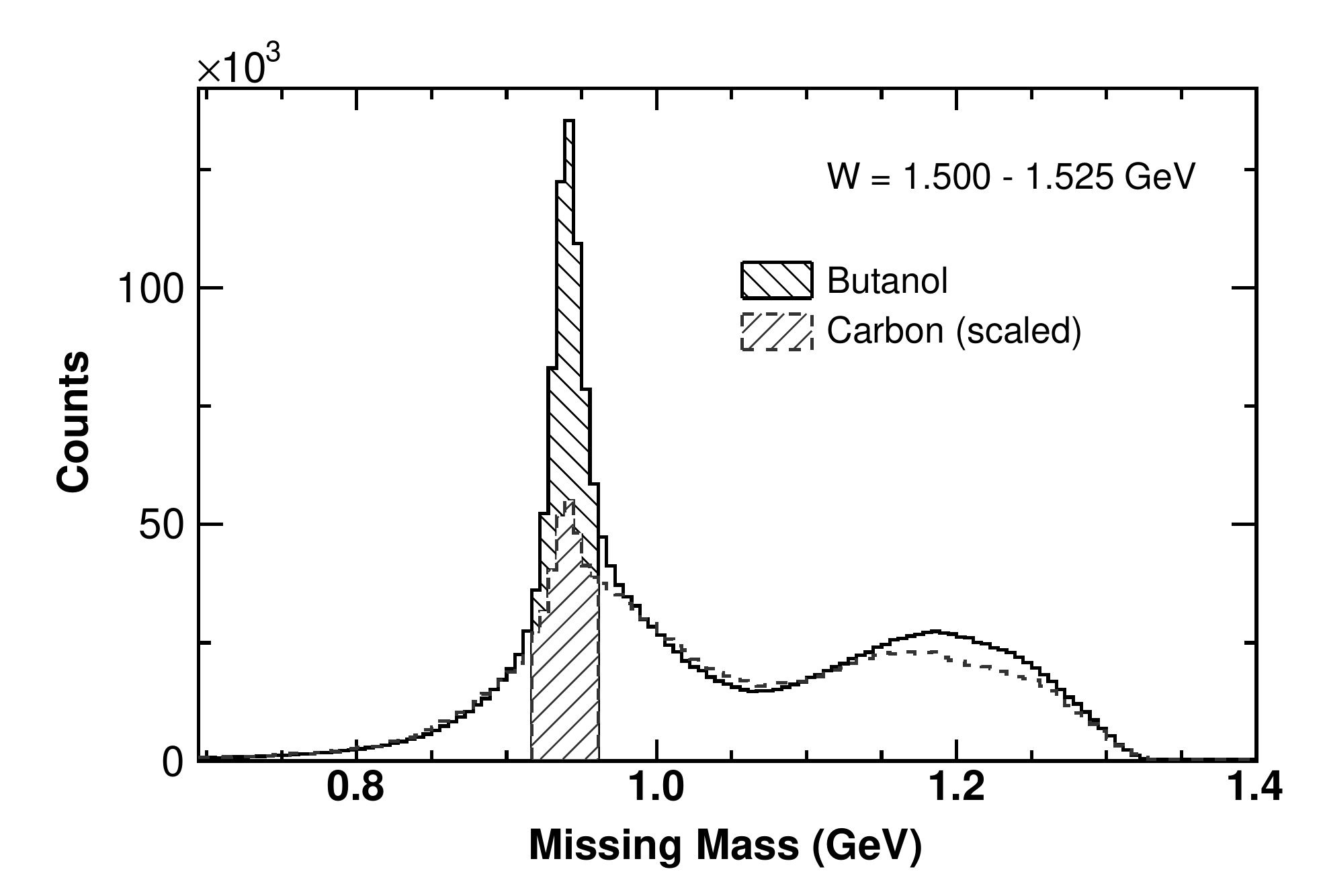}
  \caption{Missing mass distribution in the $\gamma p \rightarrow
    \pi^+X$ reaction for events originating in the butanol and
    $^{12}$C targets, respectively.  The distribution from carbon has
    been scaled to fit the butanol distribution at low values of
    missing mass. The dark hatched region identifies events of the
    $\gamma p \rightarrow \pi^+n$ reaction channel for which the
    dilution factor and the experimental raw asymmetry were
    extracted.}
  \label{fig:MM}
\end{figure}

\begin{figure}[htb!]
  \includegraphics[height=.3\textheight]{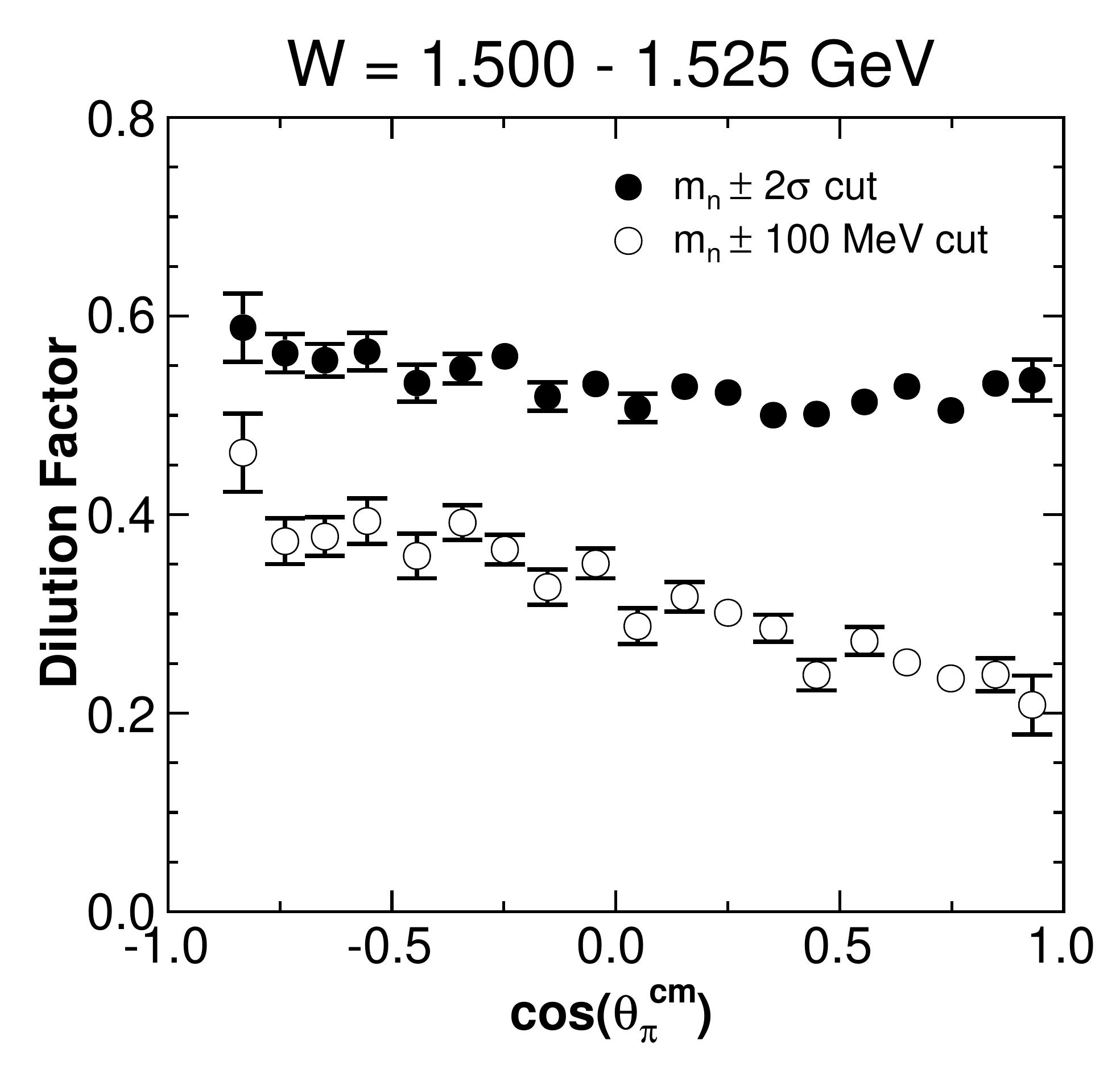}
  \caption{Angular distribution of extracted dilution factors for one
    the $W = 1.500$ to $1.525$~GeV energy bin.  The dilution factors
    are based on a selection of events with missing masses in the
    $\gamma p \rightarrow \pi^+X$ reaction close to the mass of the
    neutron; the specific cuts about the mass of the neutron are as
    indicated. }
  \label{fig:h}
\end{figure}

Figure \ref{fig:MM} shows a typical missing mass distribution for
events originating from the butanol target (solid histogram).  The
hatched region around the mass of the neutron, $m_n$, includes events
from the \pip reaction as well as events from reactions on the
unpolarized nuclei in the target, $^4$He, $^{12}$C, $^{16}$O.  All of
these events have been used in the analysis to calculate the experimental raw
asymmetry between events for which the beam spin was aligned, $N^+$,
or anti-aligned, $N^-$, with the spin of the target protons.
The butanol missing-mass distribution is compared to the distribution
of events from the $^{12}$C target (dashed histogram).  The latter
distribution was scaled to fit the butanol distribution at low missing
mass where solely events from bound nucleons contribute.  The fraction
of events originating from $^1$H relative to the total number of
events in the hatched region determines the dilution factor.  It
accounts for the suppression of the asymmetry signal due to events 
originating from unpolarized bound protons.  The dilution factor
depends on the missing-mass range used in the analysis, the kinematics
of the events, and the resolution of the detector.  This is
illustrated in Fig.~\ref{fig:h} which shows an angular distribution of
dilution factors for a given energy bin.  The dilution factors for a
more narrow missing-mass range, $\pm2\sigma$ of the neutron-mass
distribution, are larger than those for wider ranges; however, at the
slight expense of counting statistics.  Either one of them give consistent
results for $E$.

Finally, the polarization observable $E$ can be extracted for each
kinematic bin from the number of scattering events from the butanol
target, $N^\pm$, the dilution factor, $h$, and the product of beam and
target polarizations:
\begin{equation}
E =-\frac{1}{hP_zP_\odot}\frac{N^+-N^-}{N^++N^-}
\end{equation}

\begin{figure}[htb!]
  \begin{tabular}{cc}
  \includegraphics[height=.25\textheight]{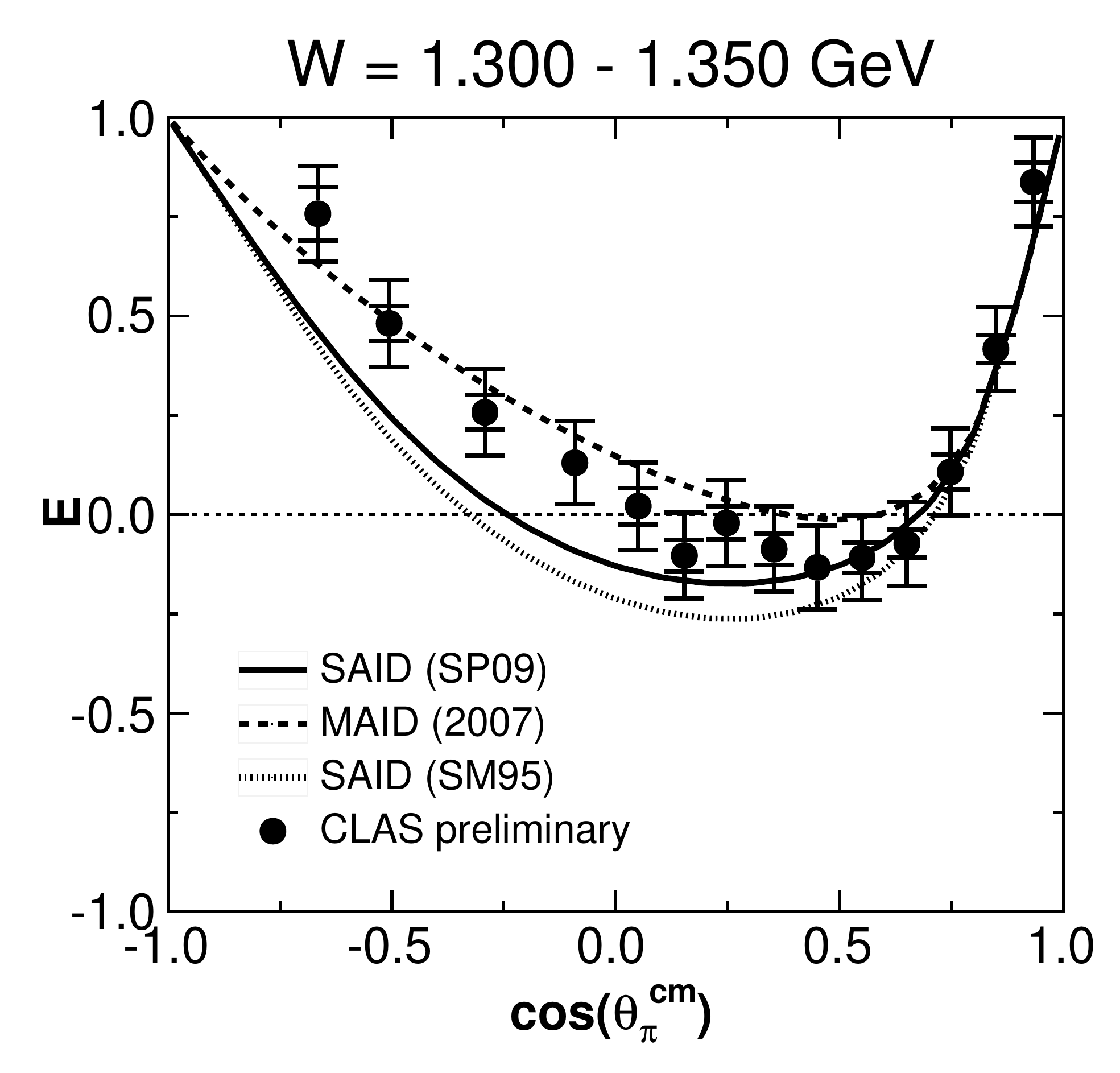} &
  \includegraphics[height=.25\textheight]{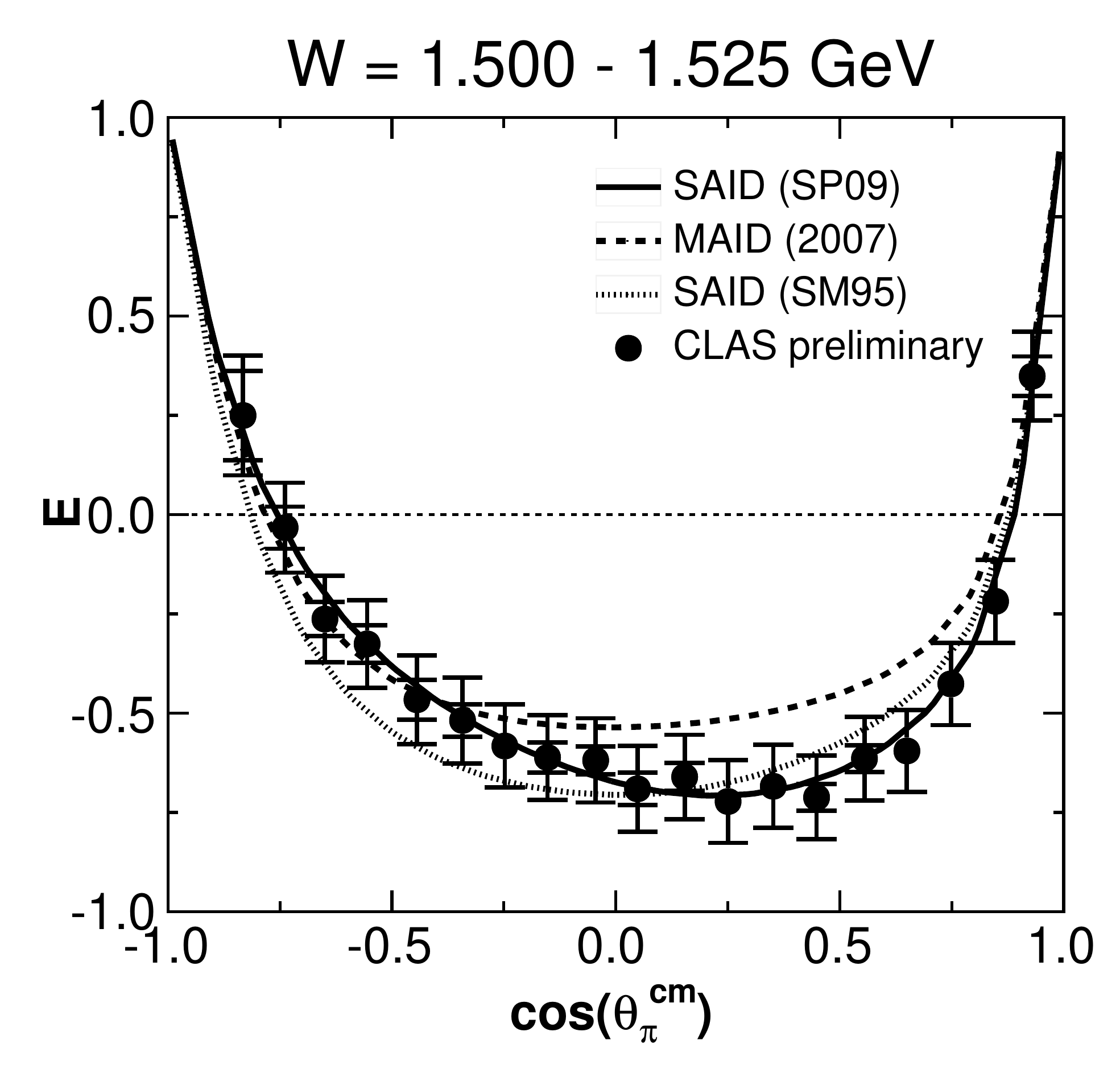} \\
  \includegraphics[height=.25\textheight]{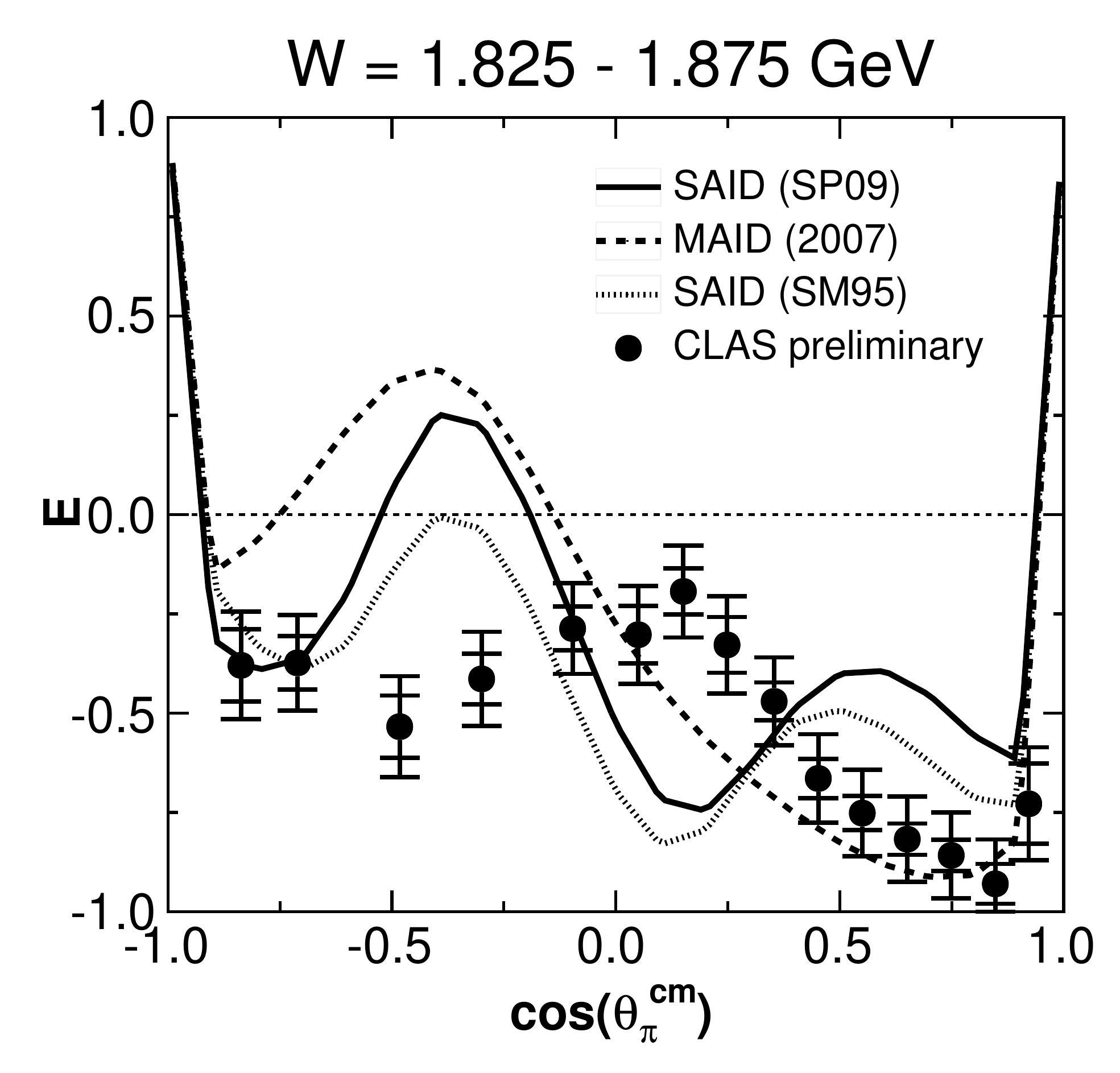} &
  \includegraphics[height=.25\textheight]{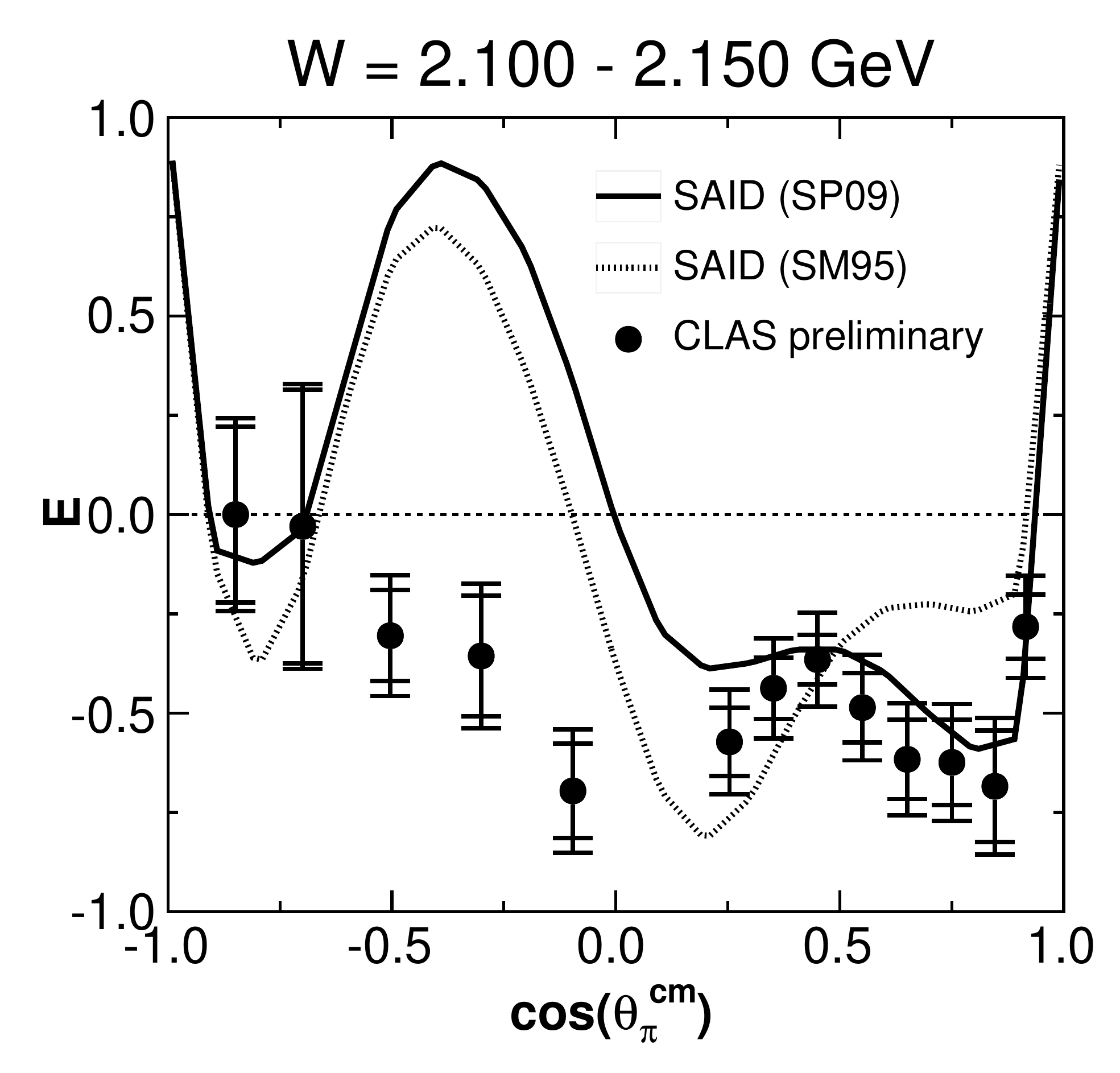}
  \end{tabular}
 \caption{Examples of preliminary results of the double-polarization
    observable $E$ in the \pip reaction for four energy bins.  The
    inner error bars indicate statistical uncertainties (not always
    visible). The outer error bars include a 10\% systematic
    uncertainty, which is expected to be reduced in the final
    analysis. The curves show present solutions of the MAID
    \cite{Drechsel:1998hk}, SAID SP09 \cite{Dugger:2009pn}, and SAID
    SM95 \cite{Arndt:1995ak} partial-wave analyses. }
  \label{fig:E}
\end{figure}

Preliminary results for $E$ are shown in Fig.~\ref{fig:E} for four
center-of-mass energy bins.  The full energy range covered in this
study is $W = 1.250$ to $2.250$ GeV in which more than 600 data points
were extracted.  The main sources of systematic uncertainties in the
experiment are from the dilution factor and the degrees of beam and
target polarizations.  The data are compared to results from MAID
\cite{Drechsel:1998hk}, SAID SP09 \cite{Dugger:2009pn}, and SAID SM95
\cite{Arndt:1995ak} partial-wave-analyses.  These models have not
been fitted to the preliminary data yet.  At energies below $W
\approx 1.7$~GeV the present solutions agree well with the main
features of the data.  As the energy increases the solutions become
less constrained and do not agree with the present data.  The
new data will greatly constrain partial-wave analyses and reduce
model-dependent uncertainties in the extraction of nucleon resonance
properties, providing a new benchmark for comparisons with
QCD-inspired models.

%


\begin{theacknowledgments}
  Work supported in parts by the U.S. National Science Foundation: NSF
  PHY-0856010. Jefferson Science Associates operates the Thomas
  Jefferson National Accelerator Facility under DOE contract
  DE-AC05-06OR23177.
\end{theacknowledgments}



\bibliographystyle{aipproc}   

\bibliography{references}

\end{document}